\documentclass[journal]{journal}

\usepackage{graphicx}
\usepackage{stmaryrd}

\begin{document}
\title{An Application of the EM-algorithm to Approximate Empirical Distributions of Financial Indices with the Gaussian Mixtures}

\author{Sergey~Tarasenko% <-this % stops a space
\thanks{S. Tarasenko is independent researcher.  Email: infra.core@gmail.com}
}
\maketitle
\thispagestyle{empty}

\begin{abstract}

In this study I briefly illustrate application of the Gaussian mixtures to approximate
empirical distributions of financial indices (DAX, Dow Jones, Nikkei, RTSI, S$\&$P 500). The resulting distributions illustrate very high quality of approximation as evaluated by Kolmogorov-Smirnov test. This implies further study of application of  the Gaussian mixtures to approximate empirical distributions of financial indices.

 \end{abstract}


\IEEEpeerreviewmaketitle

\section{Introduction}

\IEEEPARstart{A}{pproximation} of empirical distributions of financial indices using mixture of Gaussian distributinos (Gaussian mixtures) (eq. (\ref{Gm})) has been recently discussed by Tarasenko and Artukhov \cite{tararh}. Here I provide detailed explanation of steps and methods.

\begin{equation}
GM = \sum^n_{i=1}{ p_i \cdot N(\mu_i,\sigma_i)}
\label{Gm}
\end{equation}

\section{EM-algorithm for mixture separation}

\subsection{General Theory}

The effective procedure for separation of mixtures was proposed by Day \cite{day1,day2}
and Dempster et al. \cite{demps}. This procedure is based on maximization of logarithmic
likelihood function under parameters $p_1$,$p_2$,..., $p_{k-1}$,$\Theta_1$,$\Theta_2$,..., $\Theta_k$, where $k$  is number
of mixture components:

\begin{equation}
\sum_{i=1}^{n}{} ln \bigg( \sum_{j=1}^{k}{p_j \cdot f(x_i; \Theta_j)}  \bigg) \to \max_{p_j,\Theta_j}
\label{emopt}
\end{equation}

In general, the algorithms of mixture separations based on (\ref{emopt}) are called Estimation and Maximization (EM) algorithms. EM-algorithm consists of two steps: E - expectation and M-maximization. This section is focused scheme how to construct EM-algorithm.

Let $g_{ij}$ is defined as posterior porbability of oservations $x_i$ to belong to $j$-th mixture component (class):

\begin{equation}
g_{ij} = \frac{p_j \cdot f(x_i; \Theta_j)}{\sum_{j=1}^{k}{p_j \cdot f(x_i; \Theta_j)}}
\label{posterior}
\end{equation}

Posterior probability $g_{ij}$ is equal or greater then 0 and $\sum_{j=1}^{k}{g_{ij}}$ for any $i$.

Let $\Theta$ be a vector of parameters: $\Theta$ = ($p_1$,$p_2$,..., $p_{k-1}$,$\Theta_1$,$\Theta_2$,..., $\Theta_k$). Next, we deompose the logarithms likelihood function into three components:

\begin{eqnarray}
ln L(\Theta) = \sum_{i=1}^{n}{} ln \bigg( \sum_{j=1}^{k}{p_j \cdot f(x_i; \Theta_j)}  \bigg) = \\
\label{loglikeexp}
\sum_{j=1}^{k}{\sum_{i=1}^{n}{g_{ij} ln (p_j)}} \\ 
\label{c1}
+ \sum_{j=1}^{k}{\sum_{i=1}^{n}{g_{ij} ln f(x_i;\Theta_j)}} \\
\label{c2}
- \sum_{j=1}^{k}{\sum_{i=1}^{n}{g_{ij} ln (g_{ij})}}
\label{c3}
\end{eqnarray}

For this algorithm to work, the initial value $\hat{\Theta}^0$ is used to calculate inital approximations for posterior probabilities $g^0_{ij}$. This is Expectation step. Then values of $g^0_{ij}$ are used to calculate value of $\hat{\Theta}^1$ during the Maximization step.

Each of components (\ref{c1}) and (\ref{c2}) are maximized independently from each other. This is possible because component (\ref{c1}) depends only on $p_j$ ($i$=1,...,$k$), and component (\ref{c2}) depends only on $\Theta_j$ ($j$=1,...,$n$).

As a solution of optimization task (\ref{optweight})

\begin{equation}
\sum_{j=1}^{k}{\sum_{i=1}^{n}{g_{ij} ln (p_j)}} \to \max_{p_1,...,p_k}
\label{optweight}
\end{equation}

a value of $p^{(t+1)}_{j}$ for the iteration $t+1$ is calculated as:

\begin{equation}
p^{(t+1)}_{j} = \frac{1}{n} \sum_{i=1}^{n}{g^{(t)}_{ij}}
\label{weightest}
\end{equation}

where $t$ is iteration number, $t$ = 1,2, ,,,

A solution of optimization task (\ref{optdense})

\begin{equation}
\sum_{j=1}^{k}{\sum_{i=1}^{n}{g_{ij} ln f(x_i;\Theta_j)}} \to \max_{\Theta_1,...,\Theta_k}
\label{optdense}
\end{equation}

depends on a particular type of function $f(\cdot)$. 

Next we consider solution of optimization task (\ref{optdense}), when $f(\cdot)$ is Gaussian distribution.

\subsection{Mixtures of Gaussian Distributions}

Here we employ Guassian distributions:

\begin{equation}
N(x;\mu, \sigma) = \frac{1}{\sigma \sqrt{2\pi}} exp^{-\frac{(x-\mu)^2}{2\sigma^2}}
\label{guassdensity}
\end{equation}

Therefore, a specific formula to compute posterior probabilities in the case of Gaussian mixtures is

\begin{equation}
g_{ij} = \frac{exp^{-\frac{(x-\mu)^2}{2\sigma^2} + ln(p_j) - ln(\sigma_j)}}{\sum_{j=1}^{k}{exp^{-\frac{(x-\mu)^2}{2\sigma^2} + ln(p_j) - ln(\sigma_j)}}}
\label{guassposterior}
\end{equation}

According to the EM-algorithm, the task is to find value of parameters $\Theta_j$ = $(\mu_j, \sigma_j)$ by solving maximization problem (\ref{maxloglikeGauss})

\begin{eqnarray}
\sum_{j=1}^{k}{ln L_j} = \sum_{j=1}^{k}{\sum_{i=1}^{n}{ln L_j}} = \\
\sum_{j=1}^{k}{\sum_{i=1}^{n}{ln \bigg( \frac{1}{\sigma \sqrt{2\pi}} exp^{-\frac{(x-\mu)^2}{2\sigma^2}} \bigg)}} \to \max_{\Theta_1, \Theta_2, ... , \Theta_k}
\label{maxloglikeGauss}
\end{eqnarray}

The solution of this mazimization problem is given by eqs. (\ref{mean}) and (\ref{std}):

\begin{eqnarray}
\hat{\mu}_j = \frac{1}{\sum_{i=1}^{n}{g_{ij}}}\sum_{i=1}^{n}{g_{ij}x_i} \\
\label{mean}
\hat{\sigma}_j = \frac{1}{\sum_{i=1}^{n}{g_{ij}}}\sum_{i=1}^{n}{g_{ij}(x_i-\hat{\mu}_j)^2}
\label{std}
\end{eqnarray}

Having calculated optimal values for weights $p_j$ and paramaters $\Theta_j$ ($j$=1,...,$k$) during a single iteration, we apply these optimal values to obtain estimates of posterior probabilities during the Expectation step of the next iteration.

As a stop criterion, we use difference between values of loglikelyhood on iteration $t$ and iteration $t+1$:

\begin{equation}
lnL^{(t+1)}(\Theta) - lnL^{(t)}(\Theta)  < \epsilon
\end{equation}

where $\epsilon$ is infinitely small real value.

\section{An Application of the EM-algorithm to Approximate Empirical Distributions of Financial Indices with Guassian Mixtures}

In this section, I provide several examples of EM-algorithm applications to approximate empirical distributions of financial indices with Gaussian mixtures. I consider the following indices: DAX, Dow Jones Industrial, Nikkei, RTSI, and S$\&$P 500.

\begin{figure}
\centering
\includegraphics[width=9cm]{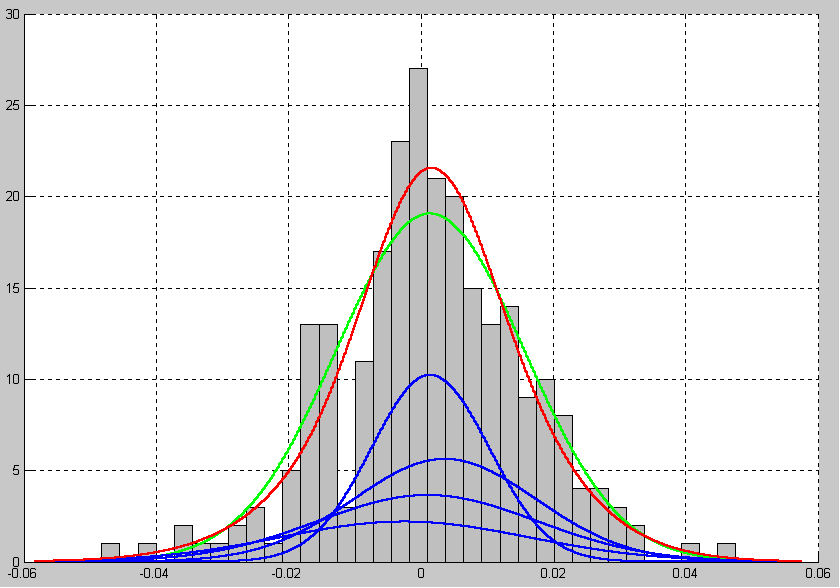}
\caption{Empirical distribution of DAX values during the period 14 April 2003 - 14 April 2004. p $<$ 0.05, KSSTAT=0.031}
\label{dax}
\end{figure}

\begin{figure}
\centering
\includegraphics[width=9cm]{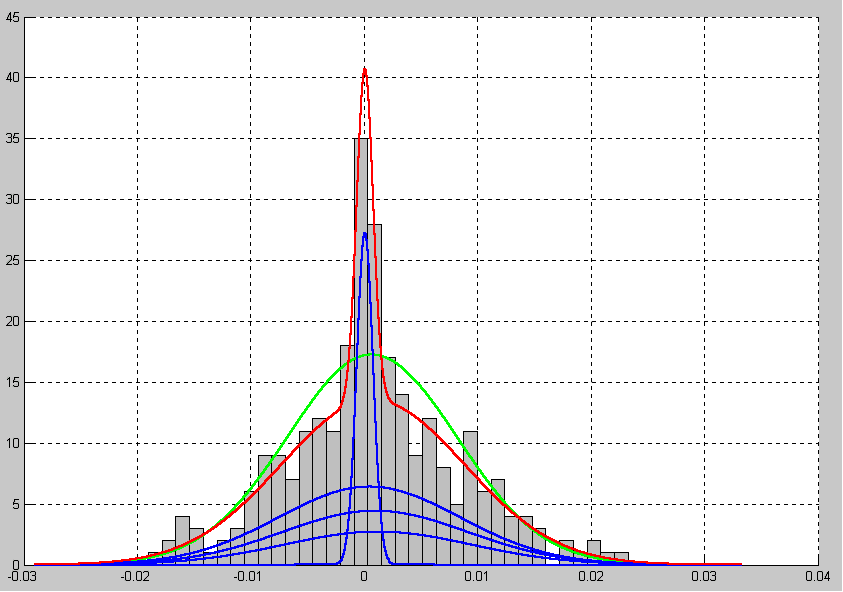}
\caption{Empirical distribution of Dow Jones Industrial values during the period 14 April 2003 - 14 April 2004, p $<$ 0.01, KSSTAT=0.023}
\label{dj}
\end{figure}

\begin{figure}
\centering
\includegraphics[width=9cm]{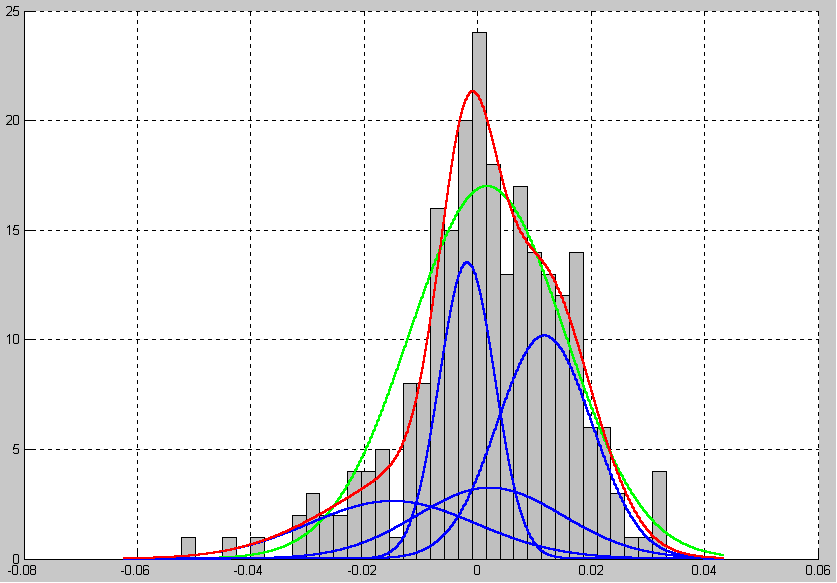}
\caption{Empirical distribution of Nikkei index values during the period 14 April 2003 - 14 April 2004, p $<$ 0.01, KSSTAT=0.027}
\label{nikkei}
\end{figure}

\begin{figure}
\centering
\includegraphics[width=9cm]{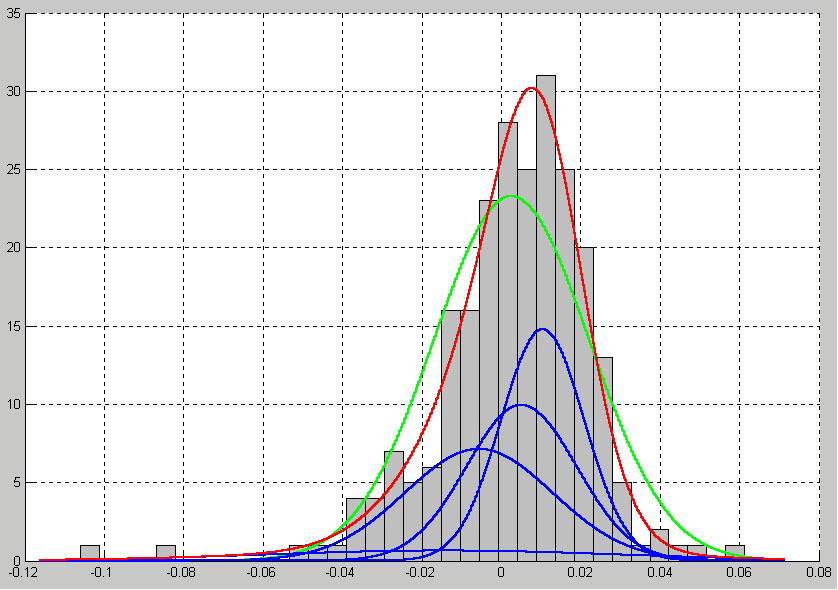}
\caption{Empirical distribution of RTSI values during the period 14 April 2003 - 14 April 2004, p $<$ 0.01, KSSTAT=0.021}
\label{rtsi}
\end{figure}

\begin{figure}
\centering
\includegraphics[width=9cm]{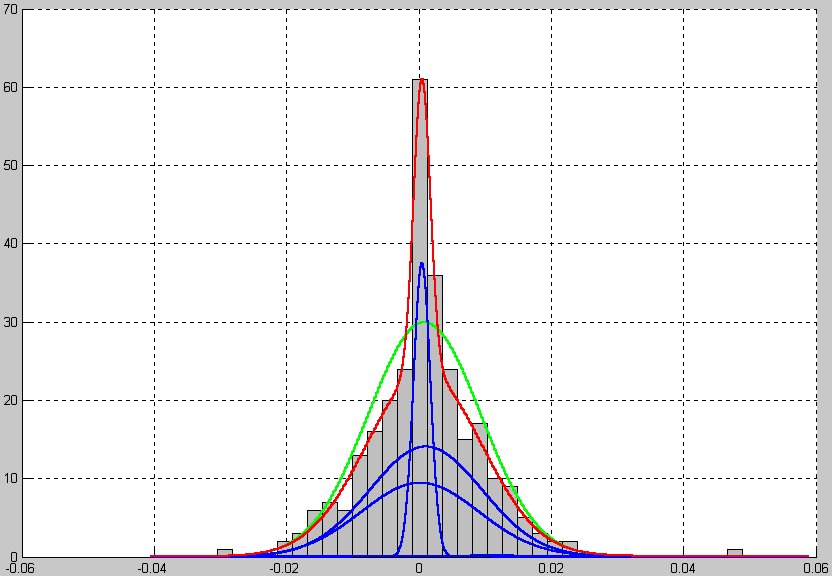}
\caption{Empirical distribution of S$\&$P 500 values during the period 14 April 2003 - 14 April 2004, p $<$ 0.01, KSSTAT=0.024}
\label{sp500}
\end{figure}

In Figs. \ref{dax}-\ref{sp500}, the green line corresponds to the Gaussian distribution, the red line illustrates a Gaussian mixture and the blue lines represent components of the Gaussian mixture.

%\textit{Analysis of DAX}

\begin{table}
\caption{Mixture model for DAX}
\centering
\begin{tabular}{c c c c} \hline
  & Weight & Mean & Standard \\
  &  & & Diviation\\ \hline
Component 1 & 0.152 & -0.002 & 0.018 \\
Component 2 & 0.223 & 0.001 & 0.017 \\
Component 3 & 0.287 & 0.004 & 0.014 \\
Component 4 & 0.337 & 0.001 & 0.009 \\ \hline
\end{tabular}
\label{daxTable}
\end{table}

%DAX: PVALUE=0.963348, KSSTAT=0.031492 \newline

%\textit{Analysis of Dow Jones}

\begin{table}
\caption{Mixture model for Dow Jones (DJIA)}
\centering
\begin{tabular}{c c c c} \hline
  & Weight & Mean & Standard \\
  &  & & Diviation\\ \hline
Component 1 & 0.173 & 0.001 & 0.008  \\
Component 2 & 0.279 & 0.001  & 0.008  \\
Component 3 & 0.396 & 0.001 & 0.008 \\
Component 4 & 0.152 & 0.000 & 0.001 \\ \hline
\end{tabular}
\label{djiaTable}
\end{table}

%DJIA: PVALUE=0.999090, KSSTAT=0.022515 \newline

%\textit{Analysis of Nikkei}

\begin{table}
\caption{Mixture model for Nikkei}
\centering
\begin{tabular}{c c c c} \hline
  & Weight & Mean & Standard \\
  &  & & Diviation\\ \hline
Component 1 & 0.167 & -0.014 & 0.014  \\
Component 2 & 0.180 & 0.002 & 0.013 \\
Component 3 & 0.367 & 0.011 & 0.008 \\
Component 4 & 0.286 & -0.002 & 0.005 \\ \hline
\end{tabular}
\label{nikkeiTable}
\end{table}

%Nikkei: PVALUE=0.994760, KSSTAT=0.026855 \newline

%\textit{Analysis of RTSI}

\begin{table}
\caption{Mixture model for RTSI}
\centering
\begin{tabular}{c c c c} \hline
  & Weight & Mean & Standard \\
  &  & & Diviation\\ \hline
Component 1 & 0.062 & -0.014 & 0.044 \\
Component 2 & 0.294 & -0.005 & 0.019 \\
Component 3 & 0.303 & 0.005 & 0.014 \\
Component 4 & 0.341 & 0.011 & 0.011 \\ \hline
\end{tabular}
\label{table1}
\end{table}

%RTSI: PVALUE=0.999885, KSSTAT=0.021401 \newline

%\textit{Analysis of S$\&$P 500}

\begin{table}
\caption{Mixture model for S$\&$P 500}
\centering
\begin{tabular}{c|c|c|c} \hline
  & Weight & Mean & Standard \\
  &  & & Diviation\\ \hline
Component 1 & 0.014 & 0.011 & 0.027 \\
Component 2 & 0.331 & 0.000 & 0.009 \\
Component 3 & 0.470 & 0.001 & 0.009 \\
Component 4 & 0.186 & 0.000 & 0.001 \\ \hline
\end{tabular}
\label{table1}
\end{table}

%SP 500: PVALUE=0.996184, KSSTAT=0.024120 \newline

\section{Discussion and Conclusion}
The results presented in this study illustrate that EM-algrothim can be effectively used to approximate empiral distributions of log daily differences of financial indices. Throughout the data for five selected indices, EM-algorithm provided very good approximation of empoiral distirbution with Gaussian mixtures. 

The approximations based on Gaussian mixtures can be used to improve application of Value-at-Risk and other methods for financial risk analysis.

This implies further explorations in applying Gaussian mixtures and the EM-algorithm for the purpose of approximation of empirial distirbutions of financial indices.


\begin{thebibliography}{1}
\bibitem{tararh} Tarasenko, S., and Artukhov, S. (2004) Stock market pricing models. In the Proceedings of International Conference of Young Scientists Lomonosov 2004, p. 248-250.

\bibitem{day1} Day, N.E. (1969) Divisive cluster analysis and test for multivariate normality. \textit{Session of the ISI}, London, 1969.

\bibitem{day2} Day, N.E. (1969) Estimating the components of a mixture of normal distributions.,
\textit{Biometrika}, 56, N3.

\bibitem{demps} Dempster, A., Laird, G. and Rubin, J. (1977) Maximum likelihood from incomplete data
via EM algorithm. \textit{Journal of Royal Statistical Society}, B, 39.

\end{thebibliography}
\end{document}